\documentclass{aa}  
\usepackage{graphicx}
\usepackage{txfonts}
\begin{document}

\title{ {\it GALEX}   colours of quasars and intergalactic medium opacity at 
    low    redshift  }
         
            \subtitle{ }

   \author{J.-M. Deharveng\inst{1}
   \and   B. Milliard\inst{1}
   \and C. P\'eroux\inst{1}
   \and  T. Small\inst{2}}

   \institute{  Aix Marseille Univ, CNRS, LAM, 
       Laboratoire d'Astrophysique de Marseille, Marseille, France
                   \email{jean-michel.deharveng@lam.fr}
    \and   1478 N.Altadena Dr, Pasadena, CA91107}


 
   \abstract
  {}
     {The distribution of neutral hydrogen  in the intergalactic medium (IGM) 
   is currently explored at low redshift by means of UV spectroscopy of quasars. We propose here an alternative approach based on UV colours of quasars as observed from GALEX surveys. We built a NUV-selected sample of 9033 quasars with (FUV$-$NUV) colours. The imprint of HI absorption in the observed colours is suggested qualitatively by their distribution as a function  of quasar redshift.}
  { Because  broad band fluxes lack spectral resolution and  are sensitive to a large range of HI column densities
   a Monte Carlo simulation of  IGM opacity is required for quantitative analysis. It was performed with absorbers randomly distributed  along redshift and column density distributions. The column density distribution was assumed to be a broken power law with index 
 $\beta_1$  ($10^{15}$  cm$^{-2} < N_{HI} <10^{17.2}$  cm$^{-2}$) and $\beta_2$ ($10^{17.2}$  cm$^{-2} < N_{HI}  <10^{19}$ cm$^{-2}$). For convenience the redshift distribution is taken proportional to the redshift evolution law of the number density of Lyman limit systems (LLS) per unit redshift as determined by existing spectroscopic surveys.The simulation is run with different assumptions 
  on the spectral index ${\alpha}_{\nu}$ of the quasar ionising flux. }
   {The fits between the simulated and observed distribution of colours require an LLS redshift density larger than that derived from spectroscopic counting. This result is robust in spite of difficulties in determining  the colour dispersion  other than that due to neutral hydrogen absorption. This difference decreases with decreasing ${\alpha}_{\nu}$  (softer ionising quasar spectrum)
 and would vanish only with values of ${\alpha}_{\nu}$ which are not supported by existing observations. }
   {We provide arguments to retain ${\alpha}_{\nu} = - 2$, a value already extreme with respect to those measured with  {\it HST}/COS.
   Further fitting of power law index  $\beta_1$ and $\beta_2$ leads to a higher density by a factor
   of 1.7 ($\beta_1 = -1.7$, $\beta_2 = -1.5$), possibly 1.5 ($\beta_1 = -1.7$,  $\beta_2 = -1.7$). Beyond the result in terms of density the analysis of UV colours of quasars reveals a tension between the current description of IGM opacity at low $z$ and the published average ionising spectrum of quasars.
}


 \titlerunning{Lyman limit systems at $0.5 < z < 1.3$ }
   \authorrunning{Deharveng et al.}

   \maketitle

%

\section{Introduction}

  Decades of quasar absorption line studies have provided a wealth of information on the  column density and redshift distributions of the neutral hydrogen pockets surviving in the nearly completely ionised intergalactic medium (IGM). These distribution functions have become basic  ingredients for the evaluation  of the IGM opacity to ionising radiation, a crucial link between the emissivity of the sources of ionising photons and the intensity of the ionising ultraviolet background radiation  
 (Haardt \& Madau 1996; Haardt \& Madau 2012).
  At large redshifts ($\ga 3$), the sources contributing to the hydrogen ionisation at a given point in the IGM are essentially local and the 
  transfer of ionising photons can be  reduced to a  mean free path term
 (Madau, Haardt  \& Rees 1999; Schirber \& Bullock 2003; Faucher-Gigu\`ere et al. 2008; Becker  \& Bolton 2013) . 
 The mean free path  
  depends on the properties of systems of optical depth near unity, usually referred as Lyman limit systems (LLSs).
  In addition, the mean free path to ionising photons can be now directly measured through composite quasar spectra 
  (Prochaska, Worseck \&  O'Meara 2009; O'Meara  et al. 2013; Fumagalli et al. 2013; Worseck   et al. 2014).

   The situation at low redshift ($ z < 2$) deserves special  attention for two reasons.
   For one,   the mean free path to ionising radiation becomes very large 
    (Fumagalli et al. 2013) and  the simplification described above is no more valid. 
Second,  the spectroscopic surveys of  low-redshift LLSs, based on the identification of the Lyman break
at 912 \AA\,
require space-based ultraviolet observations which have been  scarce. This, in combination with the intrinsically low frequency of LLSs at low redshift, explains the limited size of available samples.
Following the pioneering work of  
Tytler (1982) using   {\it IUE},     
Stengler-Larrea et al. (1995)  have homogenised the early observations with those obtained from the first spectrographs of  {\it HST}, ending with 11 LLSs 
in the redshift bin (0.36, 1.1). 
Songaila \& Cowie (2010) have added nine LLSs in the redshift bin (0.58, 1.24)
from the observations of 50 quasars with the {\it GALEX}
grism spectrograph. 
Ribaudo, Lehner \& Howk (2011) have performed an extensive analysis of archival {\it HST} quasar spectra (including from STIS), resulting  in 31 LLSs ($\tau > 1$) in their two lowest redshift bins 
(0.255, 1.060), (1.060, 1.423). More recently,
 O'Meara  et al. (2013) have analysed the incidence of LLS in 71 quasars observed with the low dispersion modes of  {\it HST}/ACS and {\it HST}/WFC3; they report 19 LLSs ($\tau > 1$) at $z < 2$ but
  with an average redshift of about 1.8 and  the bulk of  statistical power at $ 2 < z < 2.5$.

   The Lyman break  which can be identified in relatively noisy and low resolution spectra is also expected to leave a trace in the UV colours of quasars, opening the possibility to retrieve
 information on the distribution  of LLSs and more generally HI in the low redshift IGM. Samples of UV colours of quasars, as obtained from   {\it GALEX} surveys, can be 
  much larger and less biased than the spectroscopic samples described above. 
The paper is organised as follows. 
In the second section we describe how a sample  of quasars with {\it GALEX} UV colours is built and how 
 the 
presence of absorbers is  contributing the dispersion observed in the colour distribution. 
In the third section we review the challenges for extracting quantitative information from the UV colour distribution. In the fourth section we propose a simple model, including a Monte Carlo calculation of the IGM opacity to ionising photons, to  predict colour distribution from a limited set of  parameters.  In the fifth section we run the simulation  in different redshift bins, forward modelling the input parameters by comparison with observations.

\section{Hints for IGM opacity in UV colours}

  \subsection{A sample of quasars with {\it GALEX} ultraviolet colours}


      The {\it GALEX} instrument provides  a (FUV$-$NUV) colour from imaging in two ultraviolet bands, 
      the FUV (1350 - 1750 \AA\ ) and NUV (1750 - 2750 \AA\ ), with 4\farcs2 - 5\farcs3 resolution and 1\fdg25
      field of view. The two bands 
      are obtained simultaneously  using a dichroic beam splitter. The observations are performed through three types of survey, the
      {\it All-sky Imaging Survey (AIS)} with typical exposure times of 100 s,  the {\it Medium Imaging Survey (MIS)} with typical exposure times of 1500 s, the  {\it Deep Imaging Survey (DIS)} with typical exposure times of 30 ks. A full description of the instrument and the mission is given 
       by Martin et al. (2005), Morrissey et al. (2005) and Morrissey et al. (2007).



   A large sample of quasars with ultraviolet colours has been obtained by cross-correlating the 
SDSS Data Release 7 (DR7) quasar catalogue   with the 
 {\it GALEX} catalogue of UV sources, corresponding essentially to the data release GR4/GR5. The quasar catalogue 
  Schneider  et al. (2010) contains 105,783 spectroscopically confirmed quasars.
 A maximum match radius 
 of 2\farcs5 was adopted, slightly above  the recommendation of 
   Morrissey et al. (2007) based on the study of {\it GALEX} astrometry and similar to the value adopted by  Trammell  et al. (2007)
   in a previous study of the UV properties  of SDSS-selected quasars. Larger match radii have been tested and discussed   in the different context of {\it GALEX} FUV selection of high-redshift quasars by  Syphers  et al. (2009) and Worseck \& Prochaska (2011).
     Our concern here is avoiding  non quasar UV source rather than missing one line of sight 
     since a miss, provided it is not related to neutral hydrogen absorption, is not expected to introduce any selection effect.
      As we are interested by the variations of the  (FUV$-$NUV) colour that might be caused by  absorber Lyman break  within the 
 {\it GALEX} bands we have limited ourselves  to $z < 1.9$. A more conservative limit will be adopted later in the analysis for the purpose of reducing selection effects. 
In order to avoid any risk of limitation in the astrometry or the photometry at the edge of the  {\it GALEX} field of view, the useful field was limited to a radius of 0\fdg55. 
More than one UV source may be obtained within the 2\farcs5 match radius, most often in  the case 
  of quasars imaged in multiple  {\it GALEX}   exposures.  Because of the poor astrometry at low signal to noise ratio, the closest source is not necessarily the one with the highest effective exposure time in the NUV. In order to overcome this issue and to retain significant NUV flux we have applied a threshold of 900 s to the NUV exposures before selecting the nearest source. This exposure time of 900 s 
  corresponds to a clear dip in the distribution of exposure times between the    {\it AIS}  and the  {\it MIS} surveys. For homogeneity we have applied a similar cut at 800 s to the FUV exposures. This results in  a NUV-selected sample
  of 12,038 quasars  with AB magnitudes ranging from  15.7 to 24.8. 
 The (FUV$-$NUV) colours are calculated for the 9033 quasars also detected in the FUV. No correction for Galactic extinction is applied because, under current assumptions, the colour correction is only one third of the E(B$-$V) colour excess (Wyder et al. 2005).
   
\subsection{The effect of absorbers on the UV colour distribution} 

\begin{figure}
\resizebox{\hsize}{!}{\includegraphics{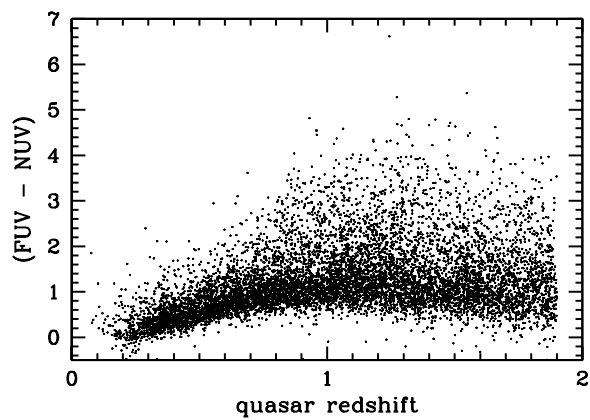}}
\caption{ (FUV$-$NUV) colours of  9033 quasars from the SDSS DR7  detected in both  {\it GALEX}  bands, as a function of their emission redshift.}
\label{fig1}
\end{figure}

\begin{figure}
\resizebox{\hsize}{!}{\includegraphics{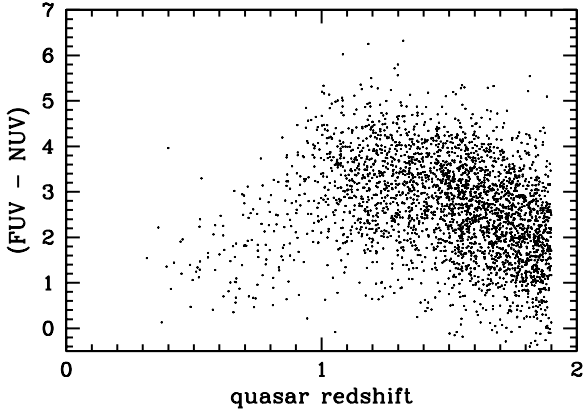}}
\caption{ Lower limit  (FUV$-$NUV) colours of  3005  quasars from the SDSS DR7  detected only in the NUV  {\it GALEX}  band, as a function of their emission redshift.}
\label{fig2}
\end{figure}

    The (FUV$-$NUV) colours of the 9033 quasars detected in both  {\it GALEX}  bands are displayed in 
  Fig.\ref{fig1} as a function of the emission redshift.  This plot is similar to one of those obtained by 
   Trammell  et al. (2007) in a previous study of  the UV properties  of SDSS-selected quasars. To deal with the 3005 quasars not detected in the FUV we have observed that the faintest FUV  detections reach an
   (AB) magnitude of 24, rather flat 
    from exposure times of 800 s to 5000 s and slowly increasing  to 25 for longer exposures. As the exposure times longer than 
    5000 s make less than 5\% of those with quasar  undetected in the FUV we have adopted 24 as a limiting magnitude to display lower limit colours of the 3005 quasars not detected in the FUV in 
    Fig.\ref{fig2}.

\begin{figure}
\resizebox{\hsize}{!}{\includegraphics{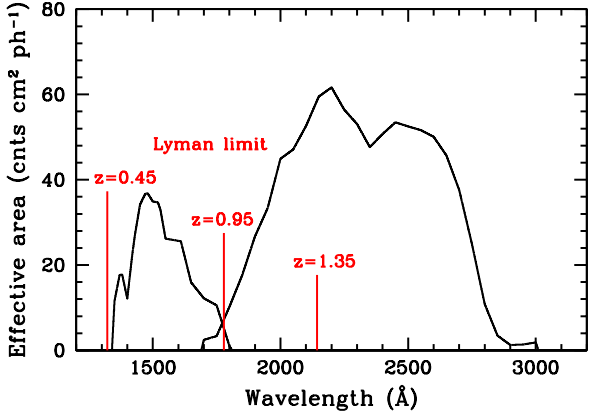}}
\caption{Effective area (counts cm$^2$ photon$^{-1}$) of the 
FUV and NUV  {\it GALEX} bandpasses (Morrissey et al. 2007).
The vertical (red) tick marks indicate the position with respect to the bandpasses of the Lyman break at redshifts 0.45, 0.95 and 1.35. }
\label{fig3}
\end{figure}

       These two first figures are best understood with 
 the details of the 
{\it GALEX}  bands  which are shown     Fig.\ref{fig3}.
     When the emission redshift $z_e$ increases from 0.45 to 0.95 the probability of a Lyman break within the FUV band increases regularly as a function of the difference $z_e - 0.45$.
     The fraction of red quasars off the main locus of points follows a similar trend in  Fig.\ref{fig1}. There is not yet any real FUV dropout in agreement  with the relatively small number of lower limit colours in  Fig.\ref{fig2}. Beyond $z_e = 0.95$ the trend flattens since an increasing  fraction of  quasars may have a Lyman break longward of the FUV band and go undetected, as seen 
   in Fig.\ref{fig2}.

\subsection{Comparison with samples of spectroscopically identified LLSs}  

     In order to support our view that the red colours of quasars may indicate the presence of neutral hydrogen absorption  we have looked at 
     the distribution of {\it GALEX}  UV colour vs. emission redshift in 
     existing samples  with spectroscopically identified LLSs at low redshift.
     
\begin{figure}
\resizebox{\hsize}{!}{\includegraphics{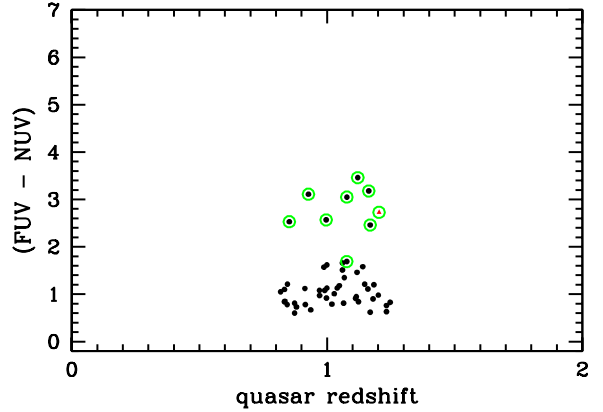}}
\caption{ (FUV$-$NUV) {\it GALEX}  colours as a function of emission redshift for the 50 quasars 
observed with {\it GALEX}  grism spectrograph by  Songaila \& Cowie (2010). 
One lower limit colour (source 
 {\it GALEX} 1712+6007)
is indicated with a red triangle. All nine quasars circled in  green colour (including the lower limit) have  spectroscopically identified LLSs. Their  LLS redshifts are all at $z < 0.95$ within the FUV band, except $z = 1.11$ for the reddest object GALEX1418+5223 which does not appear as a FUV dropout.}
\label{fig4}
\end{figure}

     We have first used the {\it GALEX} grism spectrograph observations of 50 quasars with 
     $0.8 < z_e < 1.3$ by Songaila \& Cowie (2010).
      The UV colours have been found from MAST  (Mikulski Archive for Space Telescopes)
     and are displayed in  Fig.\ref{fig4}. Of the nine quasars with a LLS spectroscopically identified by Songaila \& Cowie (2010),
      eight have colours significantly redder than the quasars without LLS, including the one with a lower limit colour which is not a FUV dropout (LLS at z = 0.82).

\begin{figure}
 \resizebox{\hsize}{!}{\includegraphics{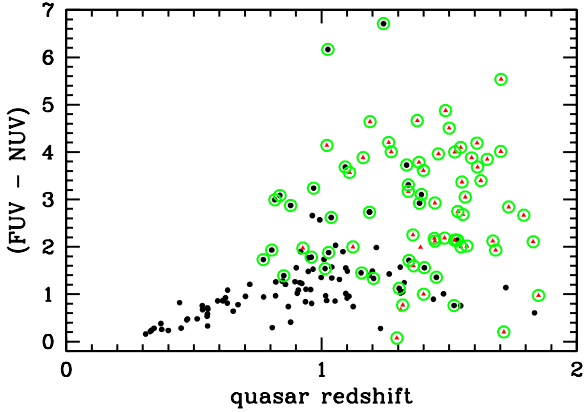}}
\caption{ (FUV$-$NUV) {\it GALEX}  colours (black dots) and lower limit colours (red triangles) as a function of emission redshift for the quasar sample used  by Ribaudo, Lehner \& Howk (2011)
for studying LLS statistics. 
Quasars circled in green colour (including the lower limits) have spectroscopically identified LLSs. The vast majority of quasars with 
(FUV$-$NUV) $> 2$ have spectroscopically identified LLSs.} 
\label{fig5}
\end{figure}

     Second, we have used the analysis by Ribaudo, Lehner \& Howk (2011) 
     of the population of LLSs at low redshift from 
     archival {\it HST} spectral observations with the Faint Object Spectrograph (FOS) and the Space Telescope Imaging Spectrograph (STIS).  The  {\it GALEX} UV colours have been searched for all the quasars with redshift $z_e < 1.85$ in their sample used for LLS statistics  (their Table 3). They are displayed in Fig.\ref{fig5}
     for   112 quasars  along with  49 quasars with only lower limit colour (no detection in the FUV). The quasars with a spectroscopically identified LLS (from their Table 4) are circled, except for 
 four of them with a Lyman break at too low redshifts to be included in the FUV band.    
     The vast majority of quasars with red colours,  that is (FUV$-$NUV)  larger than about  two, and among them a significant number of lower limits, have spectroscopically identified LLSs while those bluer have none. 
      Among  the 13 quasars circled with 
 (FUV$-$NUV) $< 2$ we have either objects classified as PLLSs (partial Lyman Limit System, $\tau < 1$) by Ribaudo, Lehner \& Howk (2011) 
 or an absorption redshift $z_{LLS}< 0.66$;  both situations come with less absorption in the {\it GALEX} FUV band.

\section{Interplay between UV colours and IGM opacity} 

          Although the distributions of  {\it GALEX} UV colours displayed in the previous sub-sections  illustrate beyond doubt the impact of LLSs (and more generally neutral hydrogen absorption) on these colours, there are many difficulties for extracting quantitative information.
          
          First of all, other factors contribute to the dispersion / reddening of UV quasar colours. In addition to photometric uncertainties, there is a natural dispersion of intrinsic quasar colours. Although the spectral energy distribution of quasars display  a relatively consistent mean successfully approximated by power laws, a significant dispersion is still present. We need to examine  how the spectral energy distributions of quasars can be described 
            in the rest-frame spectral range relevant to the {\it GALEX} colours and redshift range of interest here
             and  whether  it is possible to disentangle information on the absorbers from the uncertainties in the flux distribution.

        A related aspect is the large number of lower limit colours  coming with the quasar undetected in the FUV band. In the absence of any absorption the quasars are already red, at a typical
        (FUV$-$NUV) colour of about one. With a median NUV magnitude of 22.4 over the full sample, a number of quasars go undetected in the FUV with only a modest reddening (typically 1.6 mag associated to a FUV limiting magnitude of 25). 
        In these conditions many lower limit colours are not indicative of true FUV dropouts.

             Another  obvious difficulty in the interpretation of photometric data is the lack of spectral resolution 
     which prevents determination of any absorption redshift and  results in a  degeneracy between absorption redshift and column density. Any colour significantly redder than the average trend 
    in  Fig.\ref{fig1}
     and potentially revealing neutral hydrogen absorption may be caused by different combinations of absorption redshift and absorption strength.
This is a basic difference with a spectroscopic approach which provides a list of  well defined absorbers at specific redshifts and with a uniform optical depth limit (e.g. larger than 1 or 2) that can be translated into a distribution law of LLSs (i.e. number of LLS per unit redshift as a function of z). Here, 
the UV colour is influenced not only by the absorbers that are usually identified as LLSs (optical depth larger than 1 or HI column density $N_{HI} \geq 10^{17.2}$  cm$^{-2}$ ) but also by an increasing number of absorbers with $N_{HI} < 10^{17.2}$  cm$^{-2}$. 
In contrast to counting a few  well identified optically thick clouds, we are challenged here by  an opacity problem caused by a large population of absorbers.  
This  bears  resemblance to the direct evaluation of mean free path to ionisation radiation through the  analysis of stacked spectra by O'Meara  et al. (2013) and Worseck   et al. (2014). 

 In the same vein as the redshift vs. absorption strength degeneracy due to the lack of spectral resolution, the photometric approach is unable to distinguish absorbers 
 lying close to the quasars which are usually excluded in spectroscopic surveys (e.g. Proximate LLSs  in Prochaska et al. 2010). 
 
     Beyond $z_e = 0.95$, LLSs may appear    in the NUV band and be responsible for 
     a decrease of the quasars NUV flux depending on the HI column density of the LLS and its position in the band. The resulting sample
   bias increases rapidly with redshift and requires a cut in the redshift $z_e$ for any quantitative analysis. Some average numbers may be derived from the average redshift density of 0.82 LLS reported  in the range (1.060, 1.423) by Ribaudo, Lehner \& Howk (2011).
   For a sample of about 600 quasars as in our observations, we get about 144 LLSs in the NUV band  up to $z_e < 1.3$, with  an absorption of as much as 0.27 of the NUV flux of the quasars. Up to $z_e < 1.4$ we get about 192 LLSs with an absorption of as much as 0.4 of the NUV flux.
 A redshift cut  at $z_e < 1.35$ would therefore keep sample bias within acceptable limits for quantitative analysis.

\section{Modelling the distribution of quasar UV colours}      

     To recover quantitative information from the observed dispersion of the UV colour distribution, 
     we have generated Monte Carlo quasar spectra randomly imprinted with IGM absorption from the observed statistical properties of redshift and column density distributions of neutral hydrogen. Folded through the 
  {\it GALEX}  bandpasses, these spectra  provide a large set of colours to be compared with  the 
  observed UV colour distribution. 
  We review in the following sub-sections the physical quantities entering the calculations and expected to be constrained in this comparison.

 \subsection{Quasar spectral energy distribution}  
      
The first step of the simulation is to adopt a representative quasar spectral energy distribution over 
the rest frame wavelength range 550  \AA $ -$ 2800  \AA\ that is shifted over the  {\it GALEX}  bandpasses for quasars with  $z_e < 1.35$.
 The
   rest-frame continuum of quasars is usually described by power laws $F_{\nu}  \propto  {\nu}^ {{\alpha}_{\nu} }$ 
  or  $F_{\lambda}  \propto  {\lambda}^ {{\alpha}_{\lambda} }$   with ${\alpha}_{\nu}   =  - ({\alpha}_{\lambda}  + 2 )$  (Shull, Stevans \& Danforth (2012).
  
   At wavelengths longer than about 1100  \AA, there is a consensus on the spectral index
   which consequently  will not be retained as a parameter to be optimised through the simulation. In addition  as emission lines may play a role in the quasar colours, we have not chosen an average index but preferred to adopt the composite spectrum of   Vanden Berk et al. (2001).
   Incidentally, this composite spectrum 
 can be approximated by  a power law index of $- 0.44$ above 1200  \AA\  (Vanden Berk et al. 2001).
   In contrast, the ${\alpha}_{\nu}$ index in the 
    EUV domain ($< 912 $ \AA) should be an essential parameter of the simulation since it is in this  wavelength range that the interplay between  the quasar  ionising flux continuum 
   and the continuum absorption by  neutral hydrogen clouds is expected to change the most  the  {\it GALEX}  broadband fluxes. A number of ${\alpha}_{\nu}$ index values have already been reported 
   from UV spectroscopic observations of bright, low-z targets. 
   Zheng et al. (1997) obtain ${\alpha}_{\nu} = - 1.8$ (radio-quiet sample only)
   from {\it HST}/FOS  and Telfer et al. (2002)
   ${\alpha}_{\nu} = - 1.57 \pm 0.17 $ (radio-quiet),  $- 1.96 \pm 0.12 $ (radio-loud) from  {\it HST}/FOS/GHRS/STIS.  These  values are challenged by  ${\alpha}_{\nu} = -0.56$ from  {\it FUSE} observations (Scott et al. 2004) 
   but from {\it GALEX} spectroscopy
Barger \& Cowie (2010)   find values of $-1.76$ and $-2.01$ in good agreement with the earlier {\it HST} determinations of Zheng et al. (1997) and Telfer et al. (2002).
More recent results have not clarified the situation.
  Shull, Stevans \& Danforth (2012) and Stevans et al. (2014) report     ${\alpha}_{\nu} = -1.41  \pm  0.15 $ from  {\it HST}/COS observations.
 The first quasar stacked spectrum at $z  \sim 2.4 $ from {\it HST}/WFC3 (Lusso et al. 2015)
 has confirmed the most negative values but a value ${\alpha}_{\nu} = -0.72  \pm  0.26$, in line with that from  {\it FUSE}, has been  recently given from new {\it HST}/COS observations 
 and analysis (Tilton et al. 2016).
 The  ${\alpha}_{\nu}$ index ($< 912 $ \AA) is clearly a parameter to be optimised in our simulation.

\begin{figure}
 \resizebox{\hsize}{!}{\includegraphics{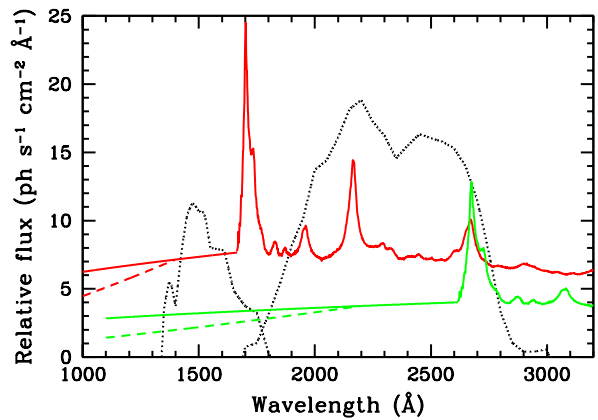}}
\caption{ Mean quasar spectral energy distribution  adopted for the simulation displayed at redshift 0.4 (red) and 1.2 (green), superposed with the  {\it GALEX} bandpasses (black dotted lines). It is composed of three parts, (i) the composite spectrum of  Vanden Berk et al. (2001)
above  1190  \AA\  rest frame,
(ii) a power law continuum below  1000  \AA\  rest frame with an index  ${\alpha}_{\nu}$ to be varied in the simulation. It is  shown here with values $-1.4$ (solid lines ${\alpha}_{\lambda} = -0.6$) and $-2.4$ (dashed lines ${\alpha}_{\lambda} = 0.4$), (iii) a power law continuum with an index $-1.4$ (${\alpha}_{\lambda} = -0.6$) between 1000  \AA\  and 1190  \AA\ (rest frame). }
\label{fig6}
\end{figure}

\begin{figure}
 \resizebox{\hsize}{!}{\includegraphics{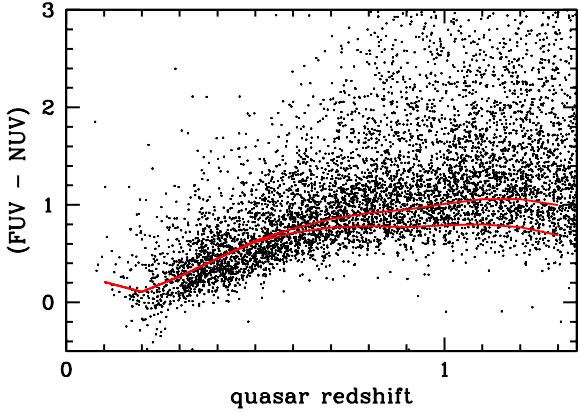}}
\caption{ Zoom-in of Fig.\ref{fig1} with the observed 
(FUV$-$NUV) {\it GALEX} colours. Superposed in red is  the colour distribution for the mean spectral energy distribution  adopted for the simulation and shown in Fig.\ref{fig6}. Two values of the power law index ${\alpha}_{\nu} $
in the wavelength range  below  1000  \AA\  rest frame have been used: $- 2.4$ (top), $-1.6$ (bottom).
Such a diagram was used to determine the index value between 1000  \AA\  and 1190  \AA\ (rest frame) 
   matching the best the observations in the range $z_e < 0.6$  (see text).}
\label{fig7}
\end{figure}

    One area of difficulty, especially for our purpose of integration into the {\it GALEX}  bandpasses, is the transition zone between the  composite spectrum of  Vanden Berk et al. (2001)
         and the softening expected in the EUV domain according to most of the values  in the references above.  A comparison with the composite spectrum of  Shull, Stevans \& Danforth (2012) 
         suggests to stop the
   composite spectrum of  Vanden Berk et al. (2001) 
   below 1190   \AA\  once the Ly$\alpha$ line has been fully covered and to start the ${\alpha}_{\nu}$ power law continuum only below
    about 1000  \AA.
      Between, we have used an intermediate  power law continuum  (Fig.\ref{fig6}). By folding the resulting complete spectral energy distribution into the  {\it GALEX} bandpasses, we have determined an index $ \sim - 1.4$
      as  best matching  the observed colour distribution of Fig.\ref{fig1}  in the domain of $z_e < 0.6$
       where the dispersion of colours is not yet affected by IGM opacity effects. This is shown in 
Fig.\ref{fig7}.  This figure illustrates how the power law index of the ionising spectrum starts to affect the quasar colours above a redshift of 0.7 and is the most sensitive in the range $0.9 - 1.3$.

\subsection{IGM absorption}  

The second step of the simulation follows the standard practice of Monte Carlo evaluation of the intergalactic hydrogen attenuation due to photoelectric absorption and resonant scattering by Lyman transitions (e.g. M{\o}ller \& Jakobsen 1990; Madau 1995; Bershady, Charlton \& Geoffroy 1999; Meiksin 2006; Inoue \& Iwata 2008; Haardt  \& Madau 2012; Inoue et al. 2014).
 At low redshift we can concentrate on the photoelectric absorption mechanism and we have not extended the simulation to the resonant scattering by Lyman transitions. To account for this latter mechanism we have used instead  the observed average transmission $ 1 -  0.013 (1 +z_e) ^{1.54} $ (observed wavelength frame)
  reported  at low redshift by Kirkman et al. (2007). 
  We have calculated that this factor has an impact of less than 0.05 magnitude on the colour in nominal conditions.
  
 We now turn to  the parameters used  in the Monte Carlo evaluation of the Lyman continuum absorption  and,  among them, the necessary limited number of those  we plan to vary and  constrain by fitting the output of the simulation with the observed quasar colour distribution. We adopt the current notation  (e.g. O'Meara  et al. 2007; Prochaska, O'Meara \& Worseck 2010)
for the column density distribution function $f (N_{HI}, X)$ giving the number of absorbers with column density $N_{HI}$ per $dN_{HI}$ interval and per absorption pathlength $dX$ (for the moment we ignore X and  the relation with the redshift). The $N_{HI}$ dependency of this function is usually approximated by broken power laws (e.g. Prochaska, O'Meara \& Worseck 2010; Ribaudo, Lehner \& Howk 2011; O'Meara  et al. 2013)
of the form 
               $ A_i N_{HI}^{\beta_i}$
               with the $A_i$ and $\beta_i$  independent of redshift and
               linked by continuity equations at the breaking points between the different classes of absorbers.

The sub-DLAs ($10^{19}$  cm$^{-2}  <  N_{HI}   <10^{20.3}$  cm$^{-2}$) and DLAs ($N_{HI}  \geq 10^{20.3}$  cm$^{-2}$) do not contribute much to the opacity because of their small numbers. As the column density distribution is also  measured  in these categories of absorbers  from the damping profiles, we have therefore adopted 
the values (respectively  $\beta_3 = - 0.8$ and $\beta_4 = - 1.4 $) reported    by    Ribaudo, Lehner \& Howk (2011)
 for the most comparable  redshift range. 
This leaves $\beta_2$ for the
 LLSs ($10^{17.2}$  cm$^{-2} <  N_{HI}     <10^{19}$  cm$^{-2}$), $\beta_1$  for the  PLLSs (partial LLS, $N_{HI}     <10^{17.2}$  cm$^{-2}$; the column density lower limit will be determined later) and the values $A_1$ (or  $A_2$)
 as parameters to optimise in our 
simulation.

\begin{figure}
 \resizebox{\hsize}{!}{\includegraphics{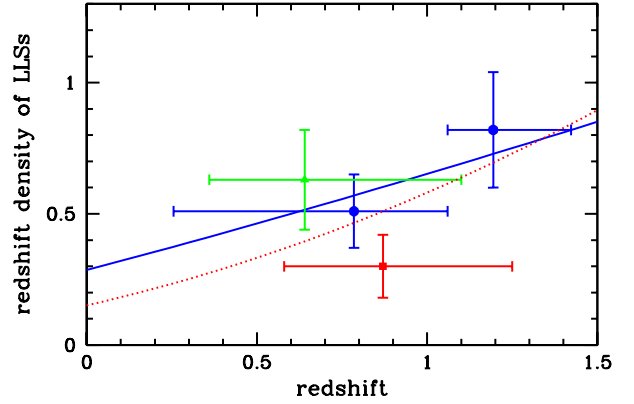}}
\caption{Previous determinations  at $ z < 1.5$ of the redshift density l(z) of LLSs ($\tau > 1$)  from spectroscopic surveys: triangle (green),   Stengler-Larrea et al. (1995)  
 with the numbers reported by Songaila \& Cowie (2010); 
 solid circle (blue), Ribaudo, Lehner \& Howk (2011);
solid square (red), Songaila \& Cowie (2010) 
from  {\it GALEX}  grism spectroscopy.
 The  lines are redshift evolution laws used for comparison: the solid line (blue) is from  Ribaudo, Lehner \& Howk (2011), 
 derived from their UV data;    the      dotted line (red) is from Songaila \& Cowie (2010), 
 based on high-redshift data and a combination of their {\it GALEX} data and those of  Stengler-Larrea et al. (1995).}
\label{fig8}
\end{figure}

     For the purpose of comparison with previous observations and in order to explore realistic values for matching the observed quasar colour distribution, it is highly desirable to establish the relation between our set of parameters   and the density of LLSs observed from previous spectroscopic surveys.
  In these surveys the observable quantity is the redshift density of LLSs $l(z)$, obtained as the ratio of 
  the number of LLSs (usually with optical depth $\tau > 1$) detected in a redshift interval to the total survey path contained in that redshift interval.
The redshift evolution law is often parameterised as  $l(z) = l_0 (1+z)^\gamma$.     
 We will refer to  those of  Ribaudo, Lehner \& Howk (2011), 
  $l(z) = 0.286 (1+z) ^{1.19}$ ($\tau > 1$) and Songaila \& Cowie (2010), 
   $l(z) = 0.151 (1+z) ^{1.94}$ displayed in  Fig.\ref{fig8} along with the density measurements from which they have been obtained.

     The quantity $l(z)$ is related to $l(X)$,
    the density of LLSs per unit pathlength X,  by e.g. Ribaudo, Lehner \& Howk (2011) and O'Meara  et al. (2013)
     $$     l(X) dX = l(z) dz   $$  where 
   $$   dX = {H_0 \over H(z)}  (1 + z)^2 dz  $$ and 
   $$   H(z) = H_0 (\Omega_{\Lambda} + \Omega_{m} (1 + z)^3)^{1/2 } $$
   
   $l(X)$ is the  zeroth moment of the column density function $f (N_{HI}, X)$ and with the broken power laws approximating this function, writes as 
   
 \begin{eqnarray}  
      l(X)_{N_{HI}>10^{17.2} cm^{-2} } 
  =  A_2 \int^ {10^{19}} _ {10^{17.2}}  N_{HI}^{\beta_2}  dN_{HI}+ \nonumber\\
      A_3 \int^ {10^{20.3}} _ {10^{19}}  N_{HI}^{-0.8}  dN_{HI}+
      A_4 \int^ {10^{22}} _ {10^{20.3}}  N_{HI}^{-1.4} dN_{HI}  \nonumber
  \end{eqnarray}

   With the relations above and $l(X)$ almost constant at low redshift, we may calculate the 
   average number of LLS absorbers ($N_{HI}  > 10^{17.2}$  cm$^{-2}$) along a line of sight 
 up to a redshift $z_e$, as (assuming  $\Omega_{\Lambda} = 0.7$ and  $\Omega_{m} = 0.3$)
     
     $$   N_{N_{HI}>10^{17.2} cm^{-2} }  =  l(X)_{N_{HI}>10^{17.2} cm^{-2} } 
     \int^ {z_e} _ {0}  {(1 + z)^2   \over  (0.7 + 0.3 (1 + z)^3)^ {1/2}}  dz $$
     
   Substituting the value of  $l(X)_{N_{HI}>10^{17.2} cm^{-2} }$  from the integration on the power laws into the last equation and comparing with the other calculation of the number of LLSs given by 
     
      $$   N_{N_{HI}>10^{17.2} cm^{-2} }   =  \int^ {z_e} _ {0}   l_0 (1+z)^\gamma dz  $$
      
      provides  a link between the set of parameters to optimise  and the LLS density measured by Songaila \& Cowie (2010) and Ribaudo, Lehner \& Howk (2011).
      This  allows us to compare the constraints from our photometric approach with the data from previous spectroscopic surveys.

   If we assume  that the shape of the redshift evolution  as a function of z used for LLSs may be extended in the domain of PLLSs, the above equations may be extended down to lower column density with the parameters A$_1$ and $\beta_1$. They may be used to determine the lowest column density $N_{HI, lim}$ with a significant role in the IGM Lyman continuum absorption.

 $$  l(X)_{N_{HI}>N_{HI, lim}} =  l(X)_{N_{HI}>10^{17.2} cm^{-2} }  +  A_1 \int^ {10^{17.2} }_ {N_{HI, lim}}  N_{HI}^{\beta_1}  dN_{HI}  $$
          and

      $$   N_{N_{HI}>N_{HI, lim}} =  l(X)_{N_{HI}>N_{HI, lim}}
     \int^ {z_e} _ {0}  {(1 + z)^2   \over  (0.7 + 0.3 (1 + z)^3)^ {1/2}}  dz $$

   We have used column density limits of $10^{14}$  cm$^{-2}$ to $10^{15}$  cm$^{-2}$ (domain of transition between the Ly-$\alpha$ forest and the PLLSs) and current values of $\beta_1$ and $\beta_2$ in the two equations above to run 
   the opacity calculation described in the following subsection. 
   Moving the limit from $10^{15}$  cm$^{-2}$ to $10^{14}$  cm$^{-2}$ would increase the total number of absorbers by a factor of about 4.9 but only  decrease the average transmitted FUV to NUV flux ratio by less than 0.002. For all practical purpose the range of column densities 
   above the limit of $10^{15}$  cm$^{-2}$ is responsible for the totality of neutral hydrogen absorption 
   and will be used in the following Monte Carlo opacity evaluation.

\subsection{Monte Carlo opacity calculation}

  For each quasar line of sight  used in our Monte Carlo opacity calculation,
   a number of absorbers  has been pseudo-randomly generated 
   from a Poisson distribution with, for mean parameter,  the number  $N_{N_{HI}>10^{15} cm^{-2} } $ obtained above for the emission redshift $z_e$. The selected number was in turn distributed pseudo-randomly on redshift and column density according to the distribution laws $l(z) = l_0 (1+z)^\gamma$ and  $ A_i N_{HI}^{\beta_i}$ in their respective 
    $ N_{HI}$ domains. 
    
    An absorber of column density $N_{HI}$ at an absorption redshift $z_a$ has an optical depth at  rest wavelength $ \lambda = \lambda_0 / (1+z_a) $ given by
    
    $ \tau =  6.3 \times 10^{-18}  {(\lambda_0  / (1+z_a) / 912)^3  }  N_{HI}$ or
    
    $ \tau = 0$     if   $ \lambda_0  / (1+z_a) > 912 \AA$
 
 The total optical depth at   $\lambda_0$ is  obtained by adding all individual absorber optical depths.   
The resulting transmission as a function of wavelength is then multiplied with the quasar spectral energy distribution (defined above with the spectral index ${\alpha}_{\nu}$) and provides a quasar spectrum with randomly imprinted IGM absorption.  Finally an integration in the   {\it GALEX}  bandpasses
gives the distribution of  quasar UV colours resulting from our Monte Carlo opacity evaluation.

\section{ The optimisation process}

       The previous section has identified the four input parameters of our Monte Carlo simulation of the distribution of  {\it GALEX}  UV colours of quasars at low redshift. They are 
 the quasar continuum spectral index ${\alpha}_{\nu}$
    in the EUV spectral range, 
    the power law index of the HI column density distribution $\beta_1$  in the range $10^{15}$  cm$^{-2} <  N_{HI}     <10^{17.2}$  cm$^{-2}$  and $\beta_2$ in the range $10^{17.2}$  cm$^{-2} <  N_{HI}   <10^{19}$  cm$^{-2}$ and 
     the density factor $A_1$.

   Given the factor larger than two between the LLS redshift densities determined so far  by spectroscopy in the far-UV range (see   Fig.\ref{fig8}) we started the optimisation with the  density factor  $A_1$ and ${\alpha}_{\nu}$ expected to affect most the simulated colour histograms.
 The density factor  $A_1$  is calculated 
  with the equations in section 4.2 and  the redshift evolution law used  for the density of LLSs.
       After a normalisation by the density factor obtained with the evolution law adopted, it is possible to work in terms of  
         a relative  number density $(1 + k)$ in each redshift bin. 
  Adopting the evolution law of Ribaudo, Lehner \& Howk (2011) 
  this means 
         that when  $k$ is 0.5 we have 1.5 times more LLSs than implied by  their distribution law.

          The parameters  $\beta_1$ and $\beta_2$ are first set to a central value  $-1.5$ and are expected to provide a possible fine tuning on the density factor  $A_1$ once determined. 
              They correspond to a range of $N_{HI}$ where most of the opacity to hydrogen-ionising photons takes place and where the column density distribution function may be difficult to establish. 
        In the range  $10^{15}$  cm$^{-2} <  N_{HI}     <10^{17.2}$  cm$^{-2}$ (index $\beta_1$)
      the absorbers are saturated in Ly$\alpha$ and, unless the Lyman break is measurable, 
       the determination of  $N_{HI}$ requires the  use of high-order Lyman series transitions  (e.g. Kim et al. 2002; Kim et al. 2013).
       Although of relatively low opacity individually, the absorbers contribute to the overall opacity through their number (e.g. Rudie et al. 2013).
    In the range $10^{17.2}$  cm$^{-2} <  N_{HI}   <10^{19}$  cm$^{-2}$  (index $\beta_2$) the HI column densities can be distinguished on the basis of the Lyman breaks only at the low range extremity and the LLS density resulting from spectroscopic survey is in fact an integral on a column density function.

 The emission redshift $z_e$ is a crucial quantity since the quasar intrinsic colour depends on the redshift and the number of absorbers
 is related  to the redshift pathlength within the  {\it GALEX}  bandpasses  whatever the random distribution of absorbers adopted. The comparison will therefore be divided into redshift intervals.
    Ten emission redshift bins  from $z_e$  = 0.4 to 1.3,  with size $z_e \pm 0.05$ and containing from 500 to 800 observed quasars are used. In each of these bins, $10^{4}$ model lines of sight are usually run for each set of parameters.
   The distribution obtained from the simulation is scaled to the number of quasars observed within the same redshift bin (including those with only lower limit on the colours). In part because of these lower limit colours, it was found more convenient to perform  the comparison and optimisation  in the form of cumulative colour histograms.
   
            \begin{figure}
 \resizebox{\hsize}{!}{\includegraphics{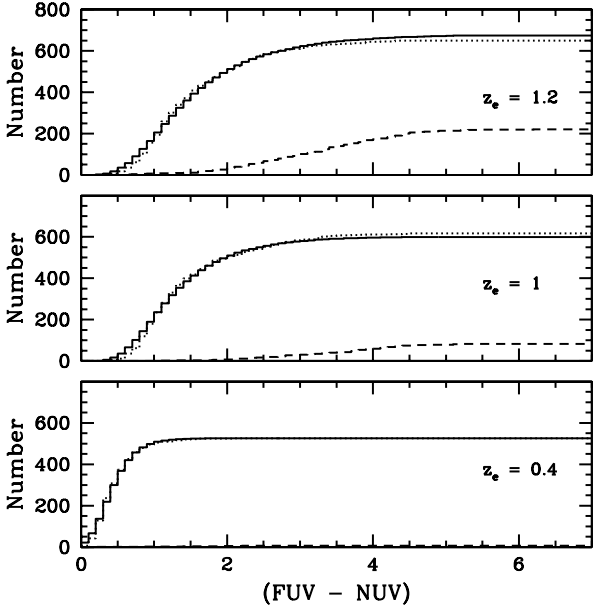}}
\caption{ Illustration of the fit between simulated (solid line) and observed (dotted line) cumulative colour histograms in different situations. 
Bottom panel: at $z_e=0.4$ (528 colour measurements; seven lower limits) the fit depends essentially on the dispersion factor and a value $\sigma = 0.05  \pm 0.005$ is determined by the best fit.
Middle panel:  fit obtained at $z_e=1.0$  with $k = 0.6$ and ${\alpha}_{\nu} = - 2.4$. 
Top panel: fit obtained at $z_e=1.2$  with $k = 0.9$ and ${\alpha}_{\nu} = - 2$.
In each panel, the dashed line is for the lower limit observed colours. Whilst it is negligible in the bottom panel, this contribution increases with redshift and remains always  larger than the number of objects with negative FUV flux produced by the simulation.
 }
  \label{fig9}
\end{figure}

   \subsection{Evaluation of a dispersion factor}    
   
   Before we can use the randomly generated distribution of quasar UV colours for comparison with the observations,  we have to add a dispersion factor to the colours that is not due to the neutral hydrogen absorption. This dispersion accounts for 
   the combination of photometric uncertainties and  a natural deviation of quasar colours from the average 
    spectral energy distribution. 
         In order to proceed with this dispersion in our simulation, we have added a normally-distributed random number with a mean zero and a standard deviation  $\sigma$ to the FUV-to-NUV flux ratio obtained for each of our quasar model line of sight. We have used the FUV-to-NUV flux ratio instead of the colour in order to account for cases with negative result and to trace them as lower limit UV colour.
    
   The low redshift bin at   $z_e  = 0.4$  where no Lyman break has yet entered the FUV {\it GALEX} band and no neutral hydrogen absorption has yet affected the colours
   is a natural place to determine this dispersion.  In the simulation the calculated  cumulative colour histogram  depends only on $\sigma$ and fits best the observations 
   for $\sigma$ = 0.05 (bottom panel of Fig.\ref{fig9}). This is consistent with the 
   rms flux ratio dispersion of 0.047 calculated with the 528 objects in  the bin $z_e = 0.4$.   
    The bin at $z_e  = 0.5$ can also be considered for the calibration of the dispersion factor since the hydrogen absorption contribution should remain marginal. The fit, not as good as for $z_e  = 0.4$,  leads to  $\sigma$ = 0.06 and gives a sense on the uncertainty on our dispersion factor.
    
\begin{figure}
 \resizebox{\hsize}{!}{\includegraphics{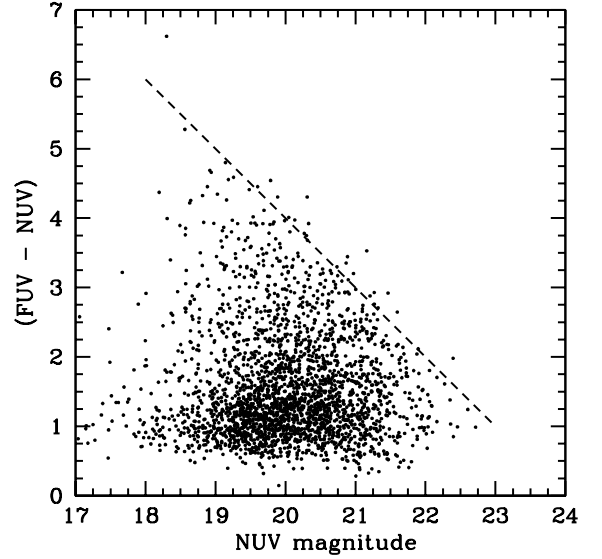}}
\caption{     (FUV $-$ NUV) colour as a function of  NUV magnitude for 2537 quasars in the range $0.95 < z_e <  1.35$.
Lower limit colours of the 688 quasars undetected in the FUV are along the short-dashed line.}
\label{fig10}
\end{figure}

 We have a number of reasons for extending up to $z_e = 1.35$ our calibration  $\sigma$ = 0.05  at low redshift. First, the NUV magnitude distribution is not changing very much among the 10 redshift bins (average magnitude per bin  between 19.76 and 20.19)   and consequently  an increase of photometric uncertainty with redshift due to UV flux getting fainter is unlikely. 
  Second, we have plotted the  observed {\it GALEX} (FUV $-$ NUV) colours as a function of their NUV magnitude in Fig.\ref{fig10}. 
  Even in the case of 
  the redshift bins 1.0 to 1.3 which exhibit the largest skewness  as seen Fig.\ref{fig1} there is no relation between the reddest and the faintest 
  quasars.
  Last, the possibility that the reddest colours may be associated to objects in high galactic extinction regions has been examined. Of the 2537 quasars in Fig.\ref{fig10} only 56 objects have  u band Galactic extinction  A$_u$ > 0.5  (Schlegel, Finkbeiner  \& Davis 1998) 
  with a maximum  A$_u$ = 1.04. 
  Among these latter 56 objects only seven have  (FUV $-$ NUV) $> 2$ while the seven with the largest A$_u$ have  (FUV $-$ NUV) $< 2$. There is no relation between the reddest colour and high galactic  extinction regions. Incidentally, 
   the extinction A$_u$ = 0.5 would 
   correspond to a shift of $-$ 0.05 and 0.842 in the  (FUV $-$ NUV) and NUV magnitude respectively
    with the corrections of Wyder et al. (2005), 
     A$_{FUV}$ = 1.584 $\times$ A$_u$ and A$_{NUV}$ = 1.684 $\times$ A$_u$.
    
      Although these series of
  arguments support our calibration of a dispersion factor at low redshift we acknowledge that we are missing an evaluation of photometric uncertainties. This is a consequence of using 
 colours instead of fluxes in our simulation, in order to stay away from 
  uncertainties  associated to UV luminosity functions of quasars.
In the following we will therefore control on purely empirical basis how each optimization may be dependent on variations 
   around the nominal dispersion factor of  $\sigma$ = 0.05

\subsection{The  bin by bin optimisation} 

 We have  explored the  fit between predicted and observed cumulative histograms up to the bin $z_e = 1.3$  with steps of 0.1 in ${\alpha}_{\nu}$ and $k$, in the ranges  $-2.6 <  {\alpha}_{\nu}  < -1.2 $ and $0 < k < 2$.  
         The steps correspond to a reasonable change that can be detected in the histogram comparison. We have proceeded  quantitatively searching the minimum of the absolute sum of the differences in each colour histogram bin up to a colour of 3.
         Fits are not considered satisfactory if this minimum is larger than about  350. 
         
          No satisfactory fit between predicted and observed cumulative histograms is found over  the redshift bins 0.6 to 0.9 as if the number of absorbers was not yet large enough nor the EUV spectral range sufficiently well shifted in the 
      {\it GALEX} bandpasses for the optimal sensitivity of the method to be reached. It is also possible that the spectral index adopted to represent the  typical quasar spectrum has  significant dispersion in the transition zone between 1190   \AA\  and 1000  \AA.
            In addition we have checked that this situation is not due to slight variations of the dispersion factor $\sigma$.

\begin{figure}
 \resizebox{\hsize}{!}{\includegraphics{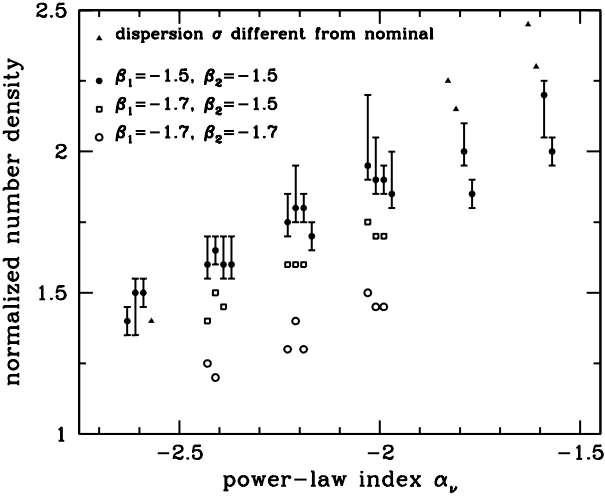}}
\caption{ Power law index  ${\alpha}_{\nu}$ and relative number density values  providing  a satisfying 
fit between the predictions and the cumulative histogram of observed colours. These are shown for 6 ${\alpha}_{\nu}$ values and 
at each of these points the data are slightly shifted in x-direction to display the 4 redshift bins, 1.0, 1.1, 1.2 and 1.3 from left to right. 
A few cases where satisfying fits are obtained  only with $\sigma $ values different from nominal are shown as solid triangles.
Open symbols show how number density values are decreased with values of $\beta_1$ and $\beta_2$ specified in the graph.}
\label{fig11}
\end{figure}

      Beyond redshift 0.9  where the neutral hydrogen absorption becomes prevalent 
     satisfactory  fits may be obtained 
       when the number  density increases (example in Fig.\ref{fig9}). This happens over a significant range of ${\alpha}_{\nu}$ values. The combinations of index ${\alpha}_{\nu}$
  and  number density (1+$k$) leading to these  fits 
  are collected and displayed in Fig.\ref{fig11}.
 Error bars corresponding to the precision on the determination of the fits are of the order of 0.05 and are shown (they were obtained by running 50,000 model lines of sight and using a step of 0.05 in $k$). These fits are obtained with the nominal value $\sigma = 0.05 $.
In some instances
a better fit may be found with different values of $\sigma  $. If the value of $k$ is different we have extended accordingly the error bar on $k$ as a sign of added uncertainty due to photometric dispersion.
Beyond the range of 
${\alpha}_{\nu}$ values   displayed, $-2.6 < {\alpha}_{\nu} < -1.5$, there is no satisfactory fits for all the redshift bins. 

  For each  set of index ${\alpha}_{\nu}$ and number density values in Fig.\ref{fig11}, we have applied the Kolmogorov-Smirnov test 
  to the two distribution functions involved in each fit and attached to  observations and simulation respectively. 
  The maximum difference between the cumulative distributions is found to vary from 0.035 to 0.105, which 
  according to the statistical test tables and the sample size of 30 bins translates into   a 95 \% significance  level  in all cases.
  
      We have finally examined how variations of the  parameters  $\beta_1$ and $\beta_2$   affect the main trend  in  Fig.\ref{fig11}. Densities can be obtained below the main trend  and are plotted  as open symbols  with indications of the relevant  $\beta_1$ and $\beta_2$ values.  The nominal value of dispersion $\sigma = 0.05 $ has been used and  error bars  similar to those in the main trend are not displayed for clarity.

 \subsection{  Results  }

A major result in 
 Fig.\ref{fig11} is  that the simulation requires larger density of LLSs than found from  previous spectroscopic determinations. 
We emphasise that 
the number density in ordinate is the fraction of absorbers in the range 
$10^{17.2}$  cm$^{-2} <  N_{HI}   <10^{19}$  cm$^{-2}$  
normalised by the density of LLSs over the same range of HI column density obtained from the  evolution law of  Ribaudo, Lehner \& Howk (2011) 
integrated to the appropriate redshift. 
Adopting the evolution law of Songaila \& Cowie (2010) 
which shows densities  lower than 
 those of  Ribaudo, Lehner \& Howk (2011) 
  in the redshift range of interest  would lead to larger overdensities.  
 Using an arbitrary evolution law is beyond the scope of our method.

The required overdensity  is dependent on 
the spectral index ${\alpha}_{\nu} $ of the quasar ionising spectrum and decreases when the index decreases and gets softer than the range of usually reported values (see sub section 4.1). This effect is stronger than the uncertainties on the dispersion $\sigma$.
The dependence on  the spectral index  does not come as a surprise since the evaluation of colours has to use an intrinsic quasar spectrum for  reference; less ionising flux from quasar should require less neutral hydrogen absorption to produce the same colour dispersion. This is illustrated by the two options of ${\alpha}_{\nu} $ plotted in 
Fig.\ref{fig7}.  At ${\alpha}_{\nu} >  -2$ we are faced to an unwanted situation with  $k$ values changing significantly with the redshift bin.
This will be discussed  in  light of all the constraints on ${\alpha}_{\nu} $ and the number density.

      Figure 11 shows also the lower overdensities values reached after optimisation of the $\beta_1$ and $\beta_2$  parameters,
       about 1.7 (${\alpha}_{\nu} = - 2$) or 1.4
   (${\alpha}_{\nu} = - 2.4 $) with $\beta_1$ decreased to about $- 1.7$.    This value of $\beta_1$ is in agreement with   
       the value $-1.65$ reported by Kim et al. (2002); Kim et al. (2003), and Rudie et al.(2013).
        Lower overdensities at about 1.5 
    (${\alpha}_{\nu} = - 2$) or 1.2
  (${\alpha}_{\nu} = - 2.4 $) may be obtained with $\beta_2$ decreased also to about $- 1.7$.   
   However, the latter value is not consistent with  a flattening 
     of the column density distribution at  $N_{HI}   > 10^{17.2}$  cm$^{-2}$ reported by  most of the observations (e.g. Prochaska, O'Meara \& Worseck 2010) and simulations (Altay et al.(2011). 

   \section{  Discussion}    

  The optimisation of the 
  power law indices  $\beta_1$ and $\beta_2$ of the HI column density distribution
  has eased but not solved the tension between the results of our approach and existing density evaluations based on spectroscopy. 
  In order to determine which combination of values to adopt from our approach  we need to examine the existing constraints on the ionising quasar spectrum (${\alpha}_{\nu}$ index) and the density of LLSs.

  \subsection{ Ionising quasar spectrum (${\alpha}_{\nu}$ index)   }   
  
  There is a large spread in the    average power  law spectral index reported for the ionising spectrum of quasars 
  from about $-2$ to $-0.56$ (see subsection 4.1).
   This range of values excludes the lowest overdensities obtained for ${\alpha}_{\nu} < -2$. 
  Constraining the overdensities within reasonable values and avoiding the unwanted dispersion with redshift observed at large ${\alpha}_{\nu}$
   in Fig.\ref{fig11} would require to stay close  to this limit ${\alpha}_{\nu} = -2$. This is in contrast with 
   the  distribution in spectral index obtained from 159 {\it HST}/COS AGN by Stevans et al. (2014) 
    with  few objects with  ${\alpha}_{\nu} < -2$.

   A possible explanation has perhaps to do with the
       duality of  the average spectral index ${\alpha}_{\nu}$. On one hand, especially in the case of simulation, the index is  for the total  ionising flux.  On the other hand the index refers only to the UV continuum of the quasar light in which case the spectral resolution is crucial for continuum placement in presence of  EUV absorption and emission lines (e.g. Stevans et al. 2014). 
       In these conditions an ${\alpha}_{\nu}$ index of about $-2$, close to determinations earlier than those taking advantage of the high spectral resolution of {\it HST}/COS, might be appropriate.

  \subsection{  LLSs density    } 
            
        While the spectroscopic determination of the density of LLSs does not include those too close to the quasar  (usually within $3000$ km s$^{-1}$), our approach coupled with opacity model does not distinguish between the ambient IGM and the proximity from quasars. This may cause an over-density in our approach. Prochaska et al. 2010 
        have discussed the incidence  of these so-called Proximate LLSs at high redshift  but we will refer to the samples of Songaila \& Cowie (2010) 
         and Ribaudo, Lehner \& Howk (2011) 
         for a comparable redshift range. Of the  nine LLSs (seven only have been kept for the density evaluation) in the 50 quasars sample of Songaila \& Cowie (2010), 
         only GALEX 1418+5223 is a potential   Proximate LLS but not fully eligible because of the proximity of detection limit and redshift uncertainty. Conservatively, we retain only an upper limit of 10 \% for the fraction of Proximate  LLS. In their larger sample of LLSs, Ribaudo, Lehner \& Howk (2011)  
         list 19 Proximate in 206 LLSs, which makes the previous limit of 10 \% a reasonable factor of overdensity due to Proximate LLSs.

\begin{figure}
 \resizebox{\hsize}{!}{\includegraphics{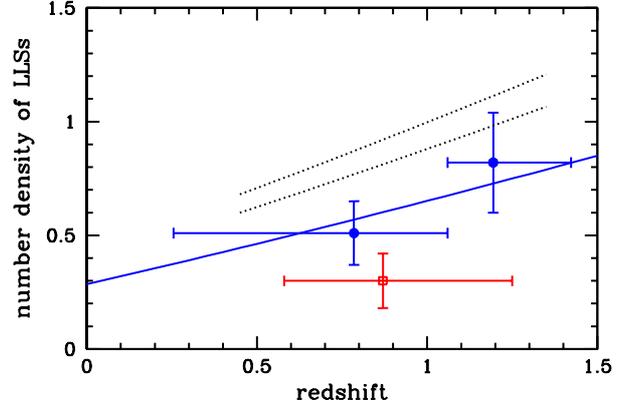}}
\caption{ Redshift density l(z) of LLSs ($\tau > 1$) per unit redshift  at $ z < 1.5$ from  Ribaudo, Lehner \& Howk (2011) 
 (blue solid  circle)
and Songaila \& Cowie (2010) 
 (red open square). Blue solid line is the redshift evolution law that Ribaudo, Lehner \& Howk (2011) 
deduced from their data and used as reference in our simulation. Overdensities of LLSs resulting from the simulation  are shown as 
black dotted lines in the range $0.45 < z < 1.35$ for  ${\alpha}_{\nu} = -2$: bottom line, over--density of 1.5 ($\beta_1 = -1.7$, $\beta_2 = -1.7$); top line, overdensity of 1.7 ($\beta_1 = -1.7$,  $\beta_2 = -1.5$). For comparison with the spectroscopic measurements these lines are shown after correction for a 10\% contribution from Proximate LLSs resulting in overdensities of 1.35 and 1.54 respectively in the plot.}
\label{fig12}
\end{figure}

      Our overdensities should be compared with the uncertainties accompanying  previous determinations from spectroscopy. 
      The measurements of Songaila \& Cowie (2010) 
      and Ribaudo, Lehner \& Howk (2011) 
      are reproduced with their error bars from Fig.\ref{fig8} into Fig.\ref{fig12}.
 The    redshift evolution law of Ribaudo, Lehner \& Howk (2011), 
 used as reference for the evaluation of overdensities, is reproduced as solid line in Fig.\ref{fig12}. 
       The overdensities are shown as dotted lines after a correction of 10\% for the contribution of Proximate LLSs. They are given with 
  ${\alpha}_{\nu} \sim -2$ and therefore considered as minimal.   
       The bottom dotted line for an over-density of 1.5  is within the one sigma error bar of the redshift density of Ribaudo, Lehner \& Howk (2011) 
        at an average $z$ of 1.19. The top dotted curve is for an overdensity 1.7 which comes with a $\beta_2$ value of $-1.5$ (instead of $-1.7$) 
       more compatible with previous determinations of $\beta_2$. 

       Finally  these overdensities have to be placed in the context of the differences between the spectroscopic determination and the  photometric approach. In the former method, the redshift density of LLSs is defined by a number of identified LLSs and a total survey path in redshift, for example 49.1 for the 31 LLSs of Ribaudo, Lehner \& Howk (2011). 
       In the latter approach, a large number of quasars is included
       over a significant redshift interval; for instance  652 quasars  with a redshift interval of about 0.7 each
       in the redshift bin $z_e = 1.2$. 
       The volume is large  but the simulation includes all types of absorbers without any specific redshift. 
       The spectroscopic counting results also in a well defined accuracy in contrast   to our approach with a precision based on the fitting process of histograms and an average   uncertainty on the overdensity curves estimated to about 0.07.
 Last, the sample of quasar colours is probably more homogeneous
 than the archival sample of Ribaudo, Lehner \& Howk (2011). 
  It is not clear that this may provide an explanation for the  overdensities since 
  an opposite effect is seen in Fig.\ref{fig12} by the measurements of Songaila \& Cowie (2010)
  with their  {\it GALEX} spectroscopic sample.

    \section { Conclusion}   
    
       We have built a sample of 9033 quasars with  {\it GALEX} (FUV $-$ NUV) colours to put constraints on the IGM neutral hydrogen absorption at low redshift (  $0.5 < z <  1.3 $ ) from  the observed dispersion of  their colours as a function of  redshift. The degeneracy between the redshift and 
 the  strength of absorption inherent to broad band fluxes  requires a Monte Carlo simulation of the IGM opacity. 
 Parameters based on existing determinations of redshift and neutral hydrogen column density distribution have been used and  a limited set of crucial parameters have been kept for the purpose of forward modelling with the observed quasar colours. They are the density of  LLSs ($10^{17.2}$  cm$^{-2} <  N_{HI}   <10^{19}$  cm$^{-2}$), the power law index $\beta_2$ of the distribution of N(HI) column density in the same range as well as the similar index $\beta_1$
 in the range
 $10^{15}$  cm$^{-2} <  N_{HI}   <10^{17.2}$  cm$^{-2}$.  
 It was convenient to normalise the  LLS density  in the simulation with the redshift density obtained by Ribaudo, Lehner \& Howk (2011) 
 from spectroscopic counting of LLSs and to adopt 
 their redshift evolution law  for running  the simulation in different redshift bins. 
 A standard quasar spectral energy distribution is also required for  
 the evaluation of quasar colour in the simulation.
 We have used existing determinations but kept  
 the power law index of the ionising spectrum ${\alpha}_{\nu} $ ($< 1000 $ \AA) as a parameter for the optimisation.
 
 The fraction of the colour dispersion not caused by neutral hydrogen absorption is calibrated at low redshift. In spite of uncertainties due to the extrapolation of this calibration at higher redshifts and the limited accuracy of the fitting process, the simulation shows a link between the derived relative LLS number density and the power law index assumed for  the ionising spectrum ${\alpha}_{\nu} $.
 Such a link is expected in the sense the more ionising photons, the more neutral hydrogen required for a given set of quasar colours.
 However, the numerical values show a tension with the existing  determinations based on spectroscopic methods.
 A number density as those obtained from these methods would come with ${\alpha}_{\nu} \sim -2.4$,  out of the range of existing 
 determinations of the ionising spectrum slope. At the opposite end a slope ${\alpha}_{\nu} \sim -1.4$, as in the last determinations 
   from  {\it HST}/COS by Shull, Stevans \& Danforth (2012), Stevans et al. (2014) but see Tilton et al. (2016),
 would come with  
   an overdensity larger than 2 and too large a dispersion between redshift bins.
   
  We  argue that a combination ${\alpha}_{\nu} = -2$ at the edge of the range of existing values of the ionising spectrum slope
 and an overdensity of 1.7 over the evolution law of redshift density of LLSs of Ribaudo, Lehner \& Howk (2011) 
 could make sense. This latter value would reduce to about 
 1.5 with  the power law index $\beta_2$ of the distribution of HI column density set to $-1.7$ instead of $-1.5$ 
 and finally to 1.35 with the correction for proximate LLSs  included. 
  Beyond the specific and modest overdensity values obtained here, this paper illustrates the potential of  large and deep sample of quasar UV colours for exploring the link between the distribution of neutral hydrogen in the IGM and  the existence of a consistent 
 mean ionising spectrum for quasars.

\begin{acknowledgements}
{C.P.  thanks the Alexander von Humboldt fundation  for the granting of a Bessel Research Award held at MPA. C.P. is also grateful to 
the  ESO  and the DFG cluster of excellence, Origin and Structure of the Universe for support. {\it GALEX (Galaxy Evolution Explorer)} is a NASA Small Explorer, launched in April 2003.
We gratefully acknowledge NASA's support for construction, operation,
and science analysis for the {\it GALEX} mission,
developed in cooperation with the Centre National d'Etudes Spatiales
of France and the Korean Ministry of 
Science and Technology. 
This research has made use of the NASA/IPAC Extragalactic Database (NED) 
which is operated by the Jet Propulsion Laboratory, California Institute of
Technology,
under contract with the National Aeronautics and Space Administration. }
\end{acknowledgements}



\end{document}